\documentclass[aps,prl,twocolumn,superscriptaddress,showpacs]{revtex4-1}

\usepackage{graphicx}
\usepackage{color}
\usepackage[normalem]{ulem}

\def\colrelax#1{\relax}

\begin{document}

\title{Soft X-ray harmonic comb from relativistic electron spikes}

\author{A.~S.~Pirozhkov}
\author{M.~Kando}
\author{T.~Zh.~Esirkepov}
\affiliation{Advanced Beam Technology Division, JAEA, 8-1-7 Umemidai, Kizugawa, Kyoto 619-0215, Japan}
\author{P.~Gallegos}
\affiliation{Central Laser Facility, Rutherford Appleton Laboratory, STFC, Chilton, Didcot, Oxon OX11 0QX, UK}
\affiliation{University of Strathclyde, Department of Physics, SUPA, Glasgow G4 0NG, UK}
\author{H.~Ahmed}
\affiliation{Centre for Plasma Physics, The Queen's University of Belfast, Belfast BT7 1NN (UK)}
\author{E.~N.~Ragozin}
\affiliation{P.~N.~Lebedev Physical Institute, RAS, Leninsky prospekt 53, Moscow 119991, Russia}
\author{A.~Ya.~Faenov}
\author{T.~A.~Pikuz}
\affiliation{Advanced Beam Technology Division, JAEA, 8-1-7 Umemidai, Kizugawa, Kyoto 619-0215, Japan}
\affiliation{Joint Institute of High Temperatures, RAS, Izhorskaja st.~13/19, Moscow 127412, Russia}
\author{T.~Kawachi}
\author{A.~Sagisaka}
\author{J.~K.~Koga}
\affiliation{Advanced Beam Technology Division, JAEA, 8-1-7 Umemidai, Kizugawa, Kyoto 619-0215, Japan}
\author{M.~Coury}
\affiliation{University of Strathclyde, Department of Physics, SUPA, Glasgow G4 0NG, UK}
\author{J.~Green}
\author{P.~Foster}
\affiliation{Central Laser Facility, Rutherford Appleton Laboratory, STFC, Chilton, Didcot, Oxon OX11 0QX, UK}
\author{C.~Brenner}
\affiliation{Central Laser Facility, Rutherford Appleton Laboratory, STFC, Chilton, Didcot, Oxon OX11 0QX, UK}
\affiliation{University of Strathclyde, Department of Physics, SUPA, Glasgow G4 0NG, UK}
\author{B.~Dromey}
\affiliation{Centre for Plasma Physics, The Queen's University of Belfast, Belfast BT7 1NN (UK)}
\author{D.~R.~Symes}
\affiliation{Central Laser Facility, Rutherford Appleton Laboratory, STFC, Chilton, Didcot, Oxon OX11 0QX, UK}
\author{M.~Mori}
\author{K.~Kawase}
\author{T.~Kameshima}
\author{Y.~Fukuda}
\affiliation{Advanced Beam Technology Division, JAEA, 8-1-7 Umemidai, Kizugawa, Kyoto 619-0215, Japan}
\author{L.~Chen}
\affiliation{Advanced Beam Technology Division, JAEA, 8-1-7 Umemidai, Kizugawa, Kyoto 619-0215, Japan}
\author{I.~Daito}
\author{K.~Ogura}
\author{Y.~Hayashi}
\author{H.~Kotaki}
\author{H.~Kiriyama}
\author{H.~Okada}
\affiliation{Advanced Beam Technology Division, JAEA, 8-1-7 Umemidai, Kizugawa, Kyoto 619-0215, Japan}
\author{N.~Nishimori}
\affiliation{Laser Application Technology Division, JAEA, 2-4 Shirakata-Shirane, Tokai, Ibaraki 319-1195, Japan}
\author{T.~Imazono}
\author{K.~Kondo}
\author{T.~Kimura}
\affiliation{Advanced Beam Technology Division, JAEA, 8-1-7 Umemidai, Kizugawa, Kyoto 619-0215, Japan}
\author{T.~Tajima}
\affiliation{Advanced Beam Technology Division, JAEA, 8-1-7 Umemidai, Kizugawa, Kyoto 619-0215, Japan}
\author{H.~Daido}
\affiliation{Advanced Beam Technology Division, JAEA, 8-1-7 Umemidai, Kizugawa, Kyoto 619-0215, Japan}
\author{P.~Rajeev}
\affiliation{Central Laser Facility, Rutherford Appleton Laboratory, STFC, Chilton, Didcot, Oxon OX11 0QX, UK}
\author{P.~McKenna}
\affiliation{University of Strathclyde, Department of Physics, SUPA, Glasgow G4 0NG, UK}
\author{M.~Borghesi}
\affiliation{Centre for Plasma Physics, The Queen's University of Belfast, Belfast BT7 1NN (UK)}
\author{D.~Neely}
\affiliation{Central Laser Facility, Rutherford Appleton Laboratory, STFC, Chilton, Didcot, Oxon OX11 0QX, UK}
\affiliation{University of Strathclyde, Department of Physics, SUPA, Glasgow G4 0NG, UK}
\author{Y.~Kato}
\affiliation{Advanced Beam Technology Division, JAEA, 8-1-7 Umemidai, Kizugawa, Kyoto 619-0215, Japan}
\author{S.~V.~Bulanov}
\affiliation{Advanced Beam Technology Division, JAEA, 8-1-7 Umemidai, Kizugawa, Kyoto 619-0215, Japan}
\affiliation{A.~M.~Prokhorov Institute of General Physics, RAS, Vavilov st.~38, Moscow 119991, Russia}

\date{\today}

\begin{abstract}

We demonstrate a new high-order harmonic generation mechanism reaching
the `water window' spectral region in experiments
with multi-terawatt femtosecond lasers irradiating gas jets.
A few hundred harmonic orders are resolved, giving $\mu$J/sr pulses.
Harmonics are collectively emitted by an oscillating electron spike
formed at the joint of the boundaries of a cavity and bow wave created by
a relativistically self-focusing laser in underdense plasma.
The spike sharpness and stability are explained by catastrophe theory.
The mechanism is corroborated by particle-in-cell simulations.
\end{abstract}


\maketitle

High-order harmonic generation,
one of the most fundamental effects of
nonlinear wave physics, originates from many nonlinearities, e.g. relativistic \cite{Teubner_Gibbon_RMP_2009_HHG,
Mourou_RMP_2006} and ionizing matter effects
\cite{McPherson_JOSAB_1987, *Corkum_PRL_1993,
*Krausz_Ivanov_RMP_2009_Atto_Phys}.
%
%
Harmonics  reaching the X-ray spectral region
enable nanometer and attosecond resolution in biology, medicine, physics,
and their applications. Several compact laser-based X-ray sources have been
implemented
, including plasma-based X-ray lasers
\cite{Matthews_PRL_1985_XRL, *Suckewer_PRL_1985_XRL, *Daido_RPP_2002_XRL},
atomic harmonics in gases \cite{McPherson_JOSAB_1987, *Corkum_PRL_1993,*Krausz_Ivanov_RMP_2009_Atto_Phys}, 
nonlinear Thomson scattering \cite{Esarey_PRE_1993_Thomson} 
from plasma electrons \cite{Chen_Nature_1998_Exp_RNTS, *Chen_PRL_2000_Phase_Matched_Thomson,*Ta_Phuoc_PRL_2003_NTS, *Banerjee_JOSAB_2003, Lee_PRE_2003_NRTS_atto} 
and electron beams 
\cite{Schoenlein_Sci_1996_90Deg_Thomson,*Babzien_PRL_2006_2_Omega_Thomson}, 
betatron radiation \cite{Albert_PPCF_2008_Full_Betatron, *Kneip_NPhys_2010_Betatron},
relativistic flying mirrors
\cite{Bulanov_PRL_2003_FM, *Kando_PRL_2007_FM, *Kando_PRL_2009_FM,
*Pirozhkov_PoP_2007_FM}, 
and harmonics from 
solid targets \cite{Mourou_RMP_2006, Teubner_Gibbon_RMP_2009_HHG, Bulanov_PoP_1994_ROM, *Dromey_NPhys_2006_HHG_Rel_Limit, *Thaury_NPhys_2007_PM_Ultrahigh_Optics, *Nomura_NPhys_2009_Atto_Phase_Lock_HHG} 
and
electro-optic shocks
\cite{Gordon_PRL_2008_Electro_Optic_Shocks}. Many of them
\cite{Esarey_PRE_1993_Thomson, Chen_Nature_1998_Exp_RNTS,
Lee_PRE_2003_NRTS_atto, Schoenlein_Sci_1996_90Deg_Thomson,
Babzien_PRL_2006_2_Omega_Thomson, Kneip_NPhys_2010_Betatron,
Teubner_Gibbon_RMP_2009_HHG, Bulanov_PoP_1994_ROM,
Dromey_NPhys_2006_HHG_Rel_Limit, Thaury_NPhys_2007_PM_Ultrahigh_Optics,
Nomura_NPhys_2009_Atto_Phase_Lock_HHG, Bulanov_PRL_2003_FM,
Kando_PRL_2007_FM, Pirozhkov_PoP_2007_FM, Kando_PRL_2009_FM} come from
relativistic laser-matter interactions, determined by the dimensionless
laser pulse amplitude ${a_0 = eE_0/m_e c\omega_0  > 1}$ related to the
laser peak irradiance by 
$I_{0L} = I_{\text{rel}} a_0^2 ({\mu\text{m}}/\lambda_0)^2$ 
for linear and 
$I_{0C} = 2 I_{0L}$ 
for circular polarization. Here, $e$ and
$m_e$ are the electron charge and mass, $c$ is the speed of light in
vacuum, $\omega_0$, $\lambda_0$ and $E_0$ are the laser angular frequency,
wavelength and peak electric field, respectively, and $I_{\text{rel}} =
1.37\times10^{18} \text{W/cm}^2$ \cite{Mourou_RMP_2006}.
%
%

In this Letter we experimentally demonstrate bright Extreme Ultraviolet (XUV) and soft X-ray harmonics resolved up to a few hundred orders from gas jets irradiated by multi-terawatt femtosecond lasers.
We propose a new mechanism of high-order harmonics generation.
%

\begin{figure}
\includegraphics[scale=0.91]{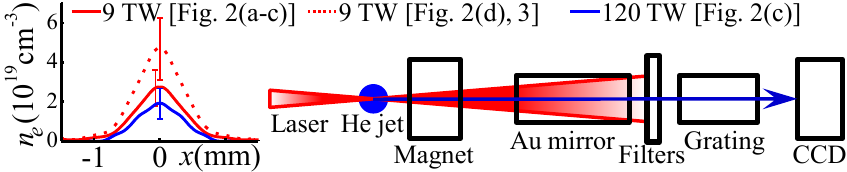}
\caption{\label{Fig_Exp_Setup}
%
(Color online) The experiment scheme (right) and He plasma density profile (left) for the shots in Figs. 2, 3.}
\end{figure}

\begin{figure}
\includegraphics[scale=0.91]{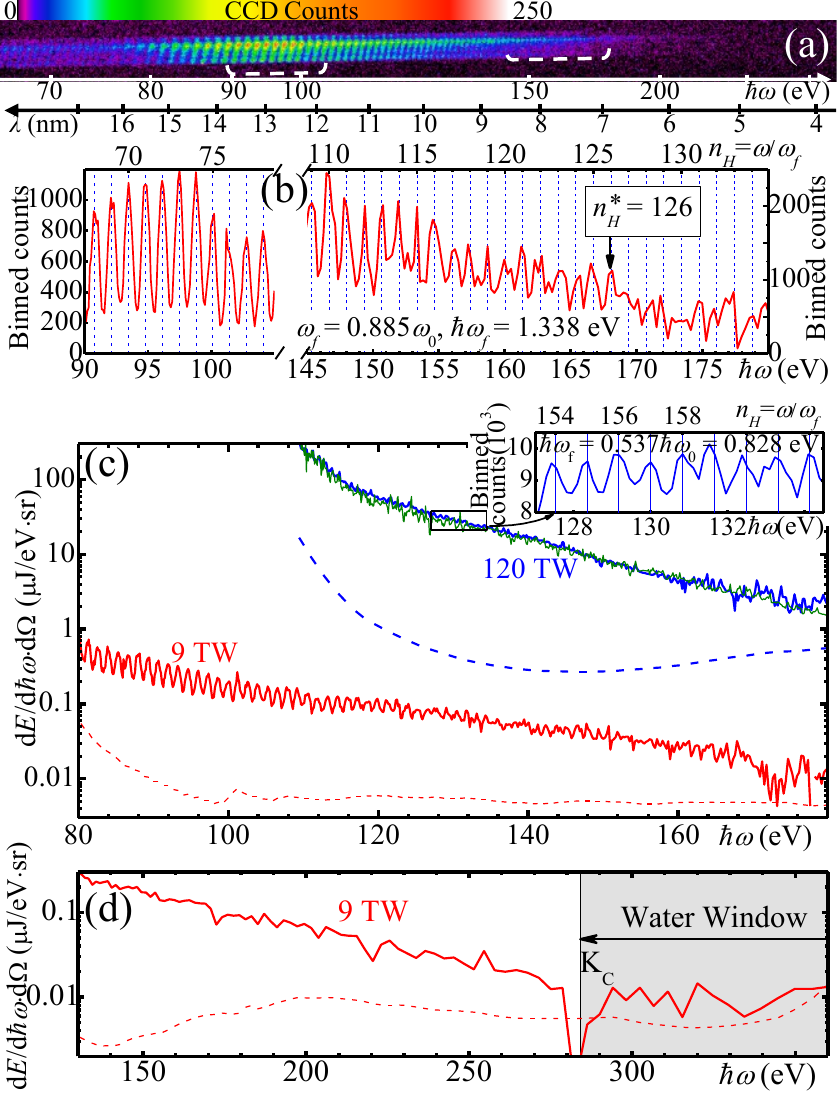}
\caption{\label{Fig_Exp_Spec}
(Color)
Typical single-shot comb-like spectra. 
(a) Raw data for $P_0=9$ TW, $n_e = 2.7\times10^{19}$ cm$^{-3}$.
(b) Lineouts for two regions marked in (a) by dashed brackets.
Dotted vertical lines for harmonics of the base frequency $\omega_f = 0.885 \omega_0$.
The highest resolved order is $n_H^*  
= 126$.
(c) Spectra for the shot shown in (a) (red) and
for $P_0=120$ TW, $n_e = 1.9\times10^{19}$ cm$^{-3}$ 
obtained from the $2^{\text{nd}}$ (blue) and $3^{\text{rd}}$ (green) diffraction orders.
Inset: 
harmonics in the 
$2^{\text{nd}}$ diffraction order; similarly for the $3^{\text{rd}}$ order.
(d) The spectral region embracing the `water window' for $P_0=9$ TW, $n_e = 4.7\times10^{19}$ cm$^{-3}$.
The drop at the carbon K absorption edge, $K_C$, is due to hydrocarbon contamination.
%
Dashed curves in (c,d) 
for background 
due to the CCD dark current, read-out noise, and scattered photons.
}
\end{figure}

\begin{figure}
\includegraphics[scale=0.91]{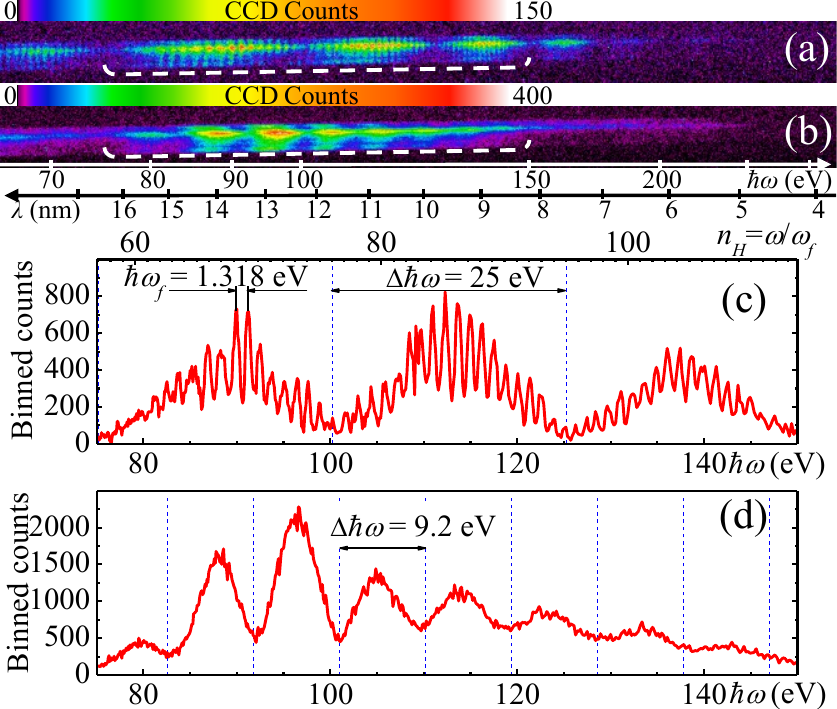}
\caption{\label{Fig_Exp_Spec_Mod}
(Color)
Modulated spectra with (a) resolved and (b) nearly unresolved
harmonic structure for $P_0=9$ TW, $n_e = 4.7\times10^{19}$ cm$^{-3}$.
Line-outs (c) and (d) show spectra denoted by dashed brackets in (a) and (b), respectively.
}
\end{figure}

We have performed two experimental campaigns using {\sf J-KAREN}
\cite{Kiriyama_OL_2008_JK} and {\sf Astra Gemini} \cite{Hooker_2006_Gemini}
lasers. The laser pulse power $P_0$, duration $\tau$, wavelength
$\lambda_0$, off-axis parabolic mirror f-number, and irradiance in vacuum are 9 TW, 27 fs, 820 nm, $f/9$, and
$4\times10^{18}$ W/cm$^2$ in the first campaign, and 120 TW, 54 fs, 804 nm, $f/20$, and $4\times10^{18}$ W/cm$^2$ in the second campaign, respectively.
Focusing laser pulses onto a supersonic helium gas jet,
Fig.~\ref{Fig_Exp_Setup},
we record harmonics in the 80-360 eV spectral region 
in the forward (laser propagation) 
direction, Figs.~\ref{Fig_Exp_Spec}, \ref{Fig_Exp_Spec_Mod}.
We use flat-field spectrographs 
\cite{Choi_AO_1997_GIS,*Neely_AIP_1998_Multichannel_GIS},
comprising a gold-coated grazing-incidence collecting mirror, 
spherical varied-linespace grating,
and back-illuminated Charge Coupling Device (CCD), shielded by
optical blocking filters -- two 0.16 $\mu$m multilayer Mo/C, two 0.2 $\mu$m
Pd, or one 0.2 $\mu$m Ag on 0.1 $\mu$m CH substrate. 
The acceptance angle is $3\times10^{-5}$ sr in the first 
and $3\times10^{-6}$ sr in the second campaign.
In the latter case, the spectrograph has two channels for
observation at 
$0^{\circ}$ (forward direction) and
$0.53^{\circ}$ in the laser (linear) polarization plane. 
The wavelength calibration is performed 
in-place using spectra of Ar and Ne plasmas. 
The harmonics energy and photon
numbers are conservatively estimated using idealized spectrograph
throughputs, i.e. the products of the calculated
\cite{Henke_1993_Atomic_Scattering_Factors} collecting mirror
reflectivities and filter transmissions, measured diffraction grating
efficiencies, and manufacturer-provided CCD efficiencies.

In both campaigns the laser power significantly exceeds the
relativistic self-focusing threshold, $P_{\rm sf} 
\approx 17$ GW$\times (n_{\text{cr}}/n_e)$, where
$n_{\text{cr}} = m \omega_0^2/4\pi e^2 \approx 1.1\times10^{21}
\text{cm}^{-3}(\mu \text{m}/\lambda_0)^2$ is 
critical density. 
Thus both lasers' pulses 
self-focus to tighter spots reaching higher irradiance determined by the plasma density and laser power. 

The comb-like spectra comprising even and odd harmonic orders of similar
intensity and shape are generated by both linearly and circularly polarized
pulses in a broad range of plasma electron densities from $n_e \simeq
1.7\times10^{19}$ to $7\times10^{19}$ cm$^{-3}$. The harmonic base frequency
$\omega_f$ is downshifted from the laser frequency $\omega_0$,
Fig.~\ref{Fig_Exp_Spec}(b,c), in correlation with the transmitted laser spectra.

%

The data obtained with different lasers demonstrate the effect's reproducibility and robustness and the photon yield scalability with the laser power, Fig.~\ref{Fig_Exp_Spec}(c).
%
%
For the 120 TW laser, 
the photon number and 
energy in unit solid angle
within one harmonic at 120 eV are
$2\times10^{12}$ photons/sr and $40\mu \text{J/sr}$, respectively. 
These amount to $
4\times 10^9$ photons and $
90$ nJ, assuming
$1.5^{\circ}$ angular radius
inferred from our particle-in-cell (PIC) simulations, 
which agrees with similar spectral intensities observed at 0 and $0.53^{\circ}$ channels.
%
Fig.~\ref{Fig_Exp_Spec}(d) 
shows 
the
spectrum 
up to the spectrograph throughput cut-off of 360 eV.
%
The emission reaches the `water window' region (284-543 eV), required for high-contrast femtosecond bioimaging,
where 
the 9 TW laser produces
$(1.5_{-0.4}^{+0.6})\times10^{10}$ photons/sr
and $0.8_{-0.2}^{+0.3}\mu \text{J/sr}$,
corresponding to 
$
3\times 10^7$ photons and 
$
1.7$ nJ for the same $1.5^{\circ}$ radius. 
The uncertainties are due to the 10\% filter thickness tolerance and CCD noise.

A large number of resolved harmonics, 
e.~g. $n_H^* \simeq 126$ in Fig.~\ref{Fig_Exp_Spec}(a,b) and $\gtrsim 160$ in
Fig.~\ref{Fig_Exp_Spec}(c, inset), strictly bounds the laser frequency
change during the harmonic emission process, $\delta \omega / \omega \leq
1/(2n_H^*)$, otherwise the orders $n_H^*$ and $n_H^*+1$ overlap.
%
%
The laser frequency decreases due to an adiabatic depletion of the laser pulse losing energy on plasma waves, while the number of photons is conserved \cite{Bulanov_Phys_Fluids_B_1992_Depletion, Esarey_RMP_2009, Shadwick_PoP_2009_Nonlinear_Depletion}.
For slow depletion,
the frequency downshift rate equals the energy depletion rate, $\delta \omega / \omega =
-\delta x/L_{\text{dep}}$ \cite{Bulanov_Phys_Fluids_B_1992_Depletion, Esarey_RMP_2009,Shadwick_PoP_2009_Nonlinear_Depletion}. 
For Fig.~\ref{Fig_Exp_Spec}(a,b), the depletion length is $L_{\text{dep}} \approx  2.7$ mm estimated from the observed $70\%$ energy transmission through the 0.9 mm plasma.
Estimate \cite{Shadwick_PoP_2009_Nonlinear_Depletion} gives 
a similar value,  $L_{\text{dep}} \approx 8.7 (n_{\text{cr}}/n_e)^{3/2}\lambda\approx 3.4$ mm.
The condition $\delta \omega / \omega < 0.4\%$ (or,
alternatively, the phase error of $<25$ mrad) gives the harmonic emission
length of $\lesssim 12 \mu \text{m}$.
We note that the emission length can be much longer than the longitudinal size of the moving source.

In many shots with linear polarization ($
40\%$ with the 9 TW laser), the
harmonic spectrum exhibits deep equidistant modulations,
Fig.~\ref{Fig_Exp_Spec_Mod}.
They are visible with discernible high orders
and even with nearly unresolved individual harmonics in some shots,
Fig.~\ref{Fig_Exp_Spec_Mod}(b,d), where blurring can be caused by
a greater downshift of the laser frequency during a longer emission of harmonics,
consistently exhibiting a few times greater photon number.
Since  in some cases the modulation depth is $\sim$100$\%$,
we conclude that  the modulations result from interference between two 
almost identical strongly localized coherent sources separated in time and/or space.
In shots with circularly polarized pulses ($90$ shots with the 120 TW laser), 
large-scale spectral modulations are not observed.


The unique properties of the observed harmonics mismatch previously suggested 
mechanisms of high-order harmonics generation.
Atomic harmonics are excluded because 
both linearly and circularly polarized laser pulses 
produce both even and odd harmonic orders
with a weak sensitivity to gas pressure.
Betatron radiation is not relevant because
the base frequency of its harmonics is determined
by the plasma frequency and electron energy, not the laser frequency.
Nonlinear Thomson scattering cannot provide the observed photon numbers
even under the most favorable assumptions.
For the 9 TW shot, Fig.~\ref{Fig_Exp_Spec}(a,b,c),
the numerical calculation of single-electron radiation spectra 
\cite{Jackson, Lee_PRE_2003_NRTS_atto}
gives at 100 eV at most
$
2\times 10^{-10}$ nJ/eV sr,
assuming
the self-focused laser amplitude of $a_0=7$, 
pulse duration of 30 fs, and the observation angle of $15^{\circ}$
close to optimum \cite{Lee_PRE_2003_NRTS_atto},
where the high-frequency nonlinear Thomson scattering component is most intense.  
The number of electrons encountered by the laser pulse
is $N_e = \pi d_{\text{sf}}^2 \Delta x n_e/4 \approx 6\times 10^9 $, 
where $n_e=2.7\times 10^{19}\text{cm}^{-3}$, 
$d_{\text{sf}} \approx 5 \mu \text{m}$ 
is the 
self-focusing channel diameter
\cite{Bulanov_SS_PoP_2010_GeV_Near_Critical}, 
and $\Delta x \approx 12 \mu \text{m}$ is the harmonics emission length 
(see above).
This gives at most 
1 nJ/eV$\cdot$sr,
200 times smaller than experimentally observed. For other spectra shown in Fig.~\ref{Fig_Exp_Spec}(c,d) even less fraction corresponds to nonlinear Thomson scattering.

\begin{figure}
\includegraphics[scale=0.85]{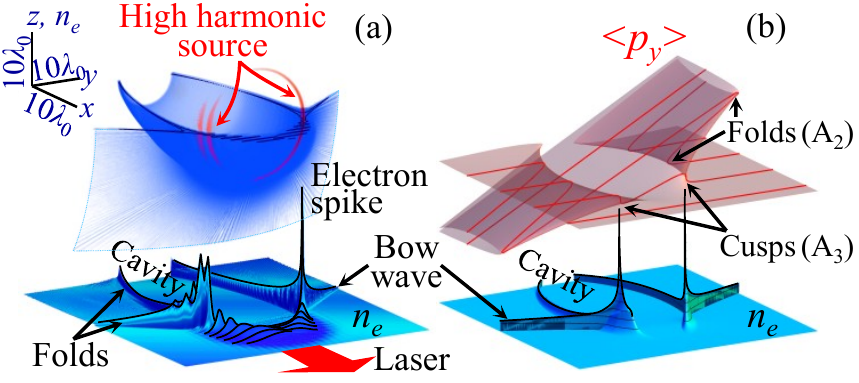}
\caption{\label{Fig_3D_PIC}
(Color)
(a) 3D PIC simulation.
Electron density $n_e$ (top, upper half removed)
and its cross-section at $z=0$ (bottom).
The electromagnetic energy density for 
$\omega \ge 4 \omega_0$ (red arcs
).
(b) Catastrophe theory model. 
Singularities in the electron density
created by foldings of the electron phase space
$(x,y,\left<p_y\right>)$, where $p_y$  is averaged over the laser period.
}
\end{figure}

\begin{figure}
\includegraphics[scale=0.85]{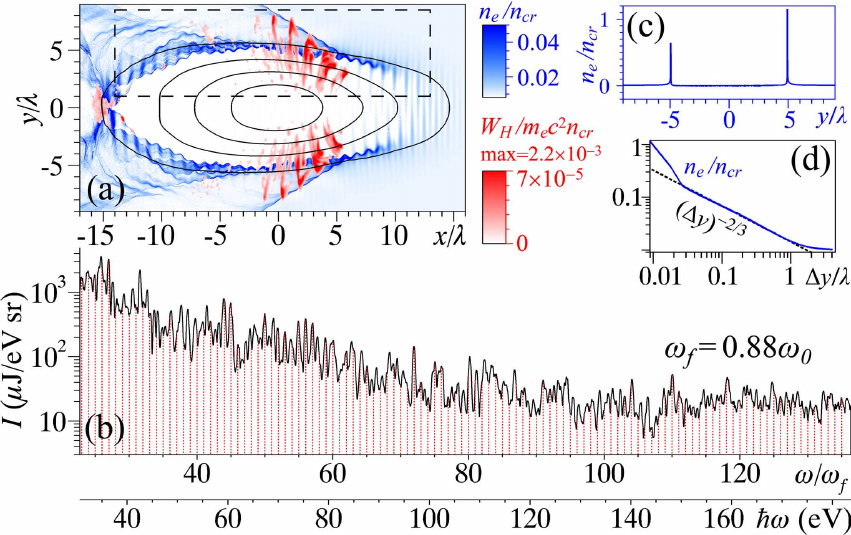}
\caption{\label{Fig_2D_PIC}
(Color)
2D PIC simulation.
(a) The electron density (blue), laser envelope (curves for $a = 1,4,7,10 $),
and electromagnetic energy density, $W_H$, for frequencies from 60 to 100$\omega_0$ (red).
(b) The upper spike emission spectrum for the dashed rectangle in (a).
(c) The electron density profile 10 laser cycles earlier than (a) and
the right spike structure (d).
}
\end{figure}

We performed 
two- (2D) and three-dimensional (3D) PIC
simulations of harmonics generation using the code {\sf REMP} \cite{Esirkepov_2001_PIC}.
As seen in simulations \cite{Suppl_material, Suppl_Movie_1}, the laser pulse undergoes self-focusing
\cite{Mourou_RMP_2006, Esarey_RMP_2009}, 
pushes electrons out evacuating a cavity in electron density
\cite{Pukhov_Bubble, Esarey_RMP_2009} and generating a bow wave
\cite{Esirkepov_PRL_2008_Bow_Wave}.
Fig.~\ref{Fig_3D_PIC}(a) represents 
the case when a linearly polarized (along $y$ axis) laser pulse 
self-focused to the amplitude of $a_0=6.6$ and FWHM waist of
$10 \lambda_0$, reaches a point where $n_e = 1.8\times 10^{18}$cm$^{-3}$.
The simulation window size is $125\lambda_0\times 124\lambda_0\times 124\lambda_0$
with the resolution of $dx=\lambda/32$, $dy=dz=\lambda/8$;
the number of quasi-particles is $2.3\times 10^{10}$; ions are immobile.
The electromagnetic energy density of a high-frequency ($\omega\ge 4\omega_0$) field
reveals the source of high-order harmonics,
Fig.~\ref{Fig_3D_PIC}(a, red arcs): they are emitted by the oscillating electron density spikes formed at the joining of the bow wave and cavity boundaries. While emission is also seen from the
cavity walls \cite{Gordon_PRL_2008_Electro_Optic_Shocks}, at higher
frequencies much stronger radiation originates from the electron spikes.

The high-order harmonics emitted by the electron spikes are seen in 2D PIC simulation, performed in the $87\lambda_0\times 72\lambda_0$ window with $dx = \lambda_0/1024$, $dy =  \lambda_0/112$ 
and
$6\times 10^8$ quasi-particles.
In Fig.~\ref{Fig_2D_PIC}, the laser pulse acquiring the amplitude of
$a_0=10$, the duration of $16\lambda_0$ and the waist of $10 \lambda_0$
meets plasma with $n_e = 1.7\times
10^{19}$cm$^{-3}$. Starting from the 7$^{\rm th}$ order, harmonics are well
discernible up to the 128$^{\rm th}$ order, 
as allowed by the simulation
resolution. The emission is slightly off-axis with the angle decreasing
with increasing harmonic order.

A strong localization of electron spikes, their robustness to oscillations
imposed by the laser and, ultimately, their superior contribution to
high-order harmonics emission is explained by catastrophe theory
\cite{Poston_Stewart}.
The laser pulse creates a multi-stream electron flow 
\cite{Bulanov_PRL_1997_Transverse_Wave_Break, Panchenko_PRE_2008_Caustics} 
stretching and folding an initially flat surface formed by electrons in their phase space, 
Fig.~\ref{Fig_3D_PIC}(b), \cite{Suppl_material, Suppl_Movie_2}.
The surface projection onto the $(x,y)$ plane gives the electron density where 
outer and inner folds are mapped into singular curves 
outlining the bow wave and cavity boundaries, respectively.
Catastrophe theory here establishes universal structurally stable
singularities, insensitive to perturbations.
The bow wave and cavity boundaries produce the `fold' type singularity (A$_2$,
according to Arnold's classification \cite{Poston_Stewart}), where the
density grows as $(\Delta y)^{-1/2}$ with decreasing distance to the
boundary, $\Delta y$.
At the joint of the two folds, the density grows as $(\Delta y)^{-2/3}$,
forming a higher order singularity -- the `cusp' (A$_3$), 
Fig.~\ref{Fig_2D_PIC}(c,d), \cite{Suppl_material, Suppl_Movie_3}.
Stronger singularities exist \cite{Panchenko_PRE_2008_Caustics}, 
however they are not stable against perturbations.
Located in a ring surrounding the cavity head, the cusp is seen in
simulations as an electron spike. 
For linear polarization, the harmonic emitting ring breaks up into two
segments, Fig.~\ref{Fig_3D_PIC}(a).
%
%
%
The spike oscillations imposed by the laser generate high-order harmonics.
The cusp singularity ensures a tight concentration of electric charge, 
making the emission coherent, 
i.~e. the intensity is proportional to the particle number squared $N_e^2$, 
similarly to the coherent nonlinear Thomson scattering. 
However, the cusp consists of different particles at every moment of time, in contrast to a synchronous motion of the same particles.
We note that for constructive interference
it is sufficient that the source size is smaller than the emitted wavelength in the direction of observation only.
The estimated number of electrons within the singularity ring, $N_e \sim
10^6$, provides a signal level close to the experiment.

In classical electrodynamics \cite{Jackson}, an oscillating electron emits harmonics up to critical order
$n_{Hc}$, proportional to the cube of the particle energy ${\cal E}_e
\approx a_0 mc^2$,
\begin{equation}\label{eq:1}
n_{Hc} = \omega_c/\omega_f \sim a_0^3,
\end{equation}
then the spectrum exponentially vanishes as seen in our experiments. This
gives clues about the harmonics generation scaling. For the laser
dimensionless amplitude in the stationary self-focusing
\cite{Bulanov_SS_PoP_2010_GeV_Near_Critical}, 
$a_{0\text{,sf}} = (8\pi P_0 n_e/P_c n_{\text{cr}})^{1/3}$, 
we obtain the critical harmonic order of
$n_{Hc} \sim P_0 n_e/P_c n_{\text{cr}}$, 
where $P_c = 2m_e^2c^5/e^2 \approx 17$ GW. 
The total energy ${\cal E}_s$ emitted by the electron spike \cite{Jackson} is
proportional to 
$N_e^2$  and the harmonic emission time $\tau_H$:
\begin{equation}\label{eq:2}
{\cal E}_s \approx e^2 N_e^2 a_0^4 \gamma \omega_0^2 \tau_H/8c \propto 
N_e^2 P_0^{4/3} n_e^{5/6} \omega_0^{1/3} \tau_H .
\end{equation}
Here $\gamma \approx (n_{cr}/n_e)^{1/2}$ is the Lorentz factor associated with the spike velocity, which is close to the laser pulse group velocity in plasma. In the 120 TW shot,
Fig.~\ref{Fig_Exp_Spec}(c), the energy emitted into a harmonic near 120 eV
is 
40 $\mu$J/sr. In the 2D PIC simulation with the parameters close
to this shot (Fig.~\ref{Fig_2D_PIC}), the energy of 100$^{\rm th}$ harmonic is 
30 $\mu$J/sr.
An estimate based on Eq.~(\ref{eq:2}) gives 100 $\mu$J/sr within the 100$^{\rm th}$
harmonic, thus providing a simple analytical estimate for the expected harmonic energy. 

With the detector at 1.4 m and the source size of 10 $\mu$m estimated from simulations,
the spatial coherence width is 1 mm, which is large enough for many phase contrast imaging applications in a compact setup
\cite{Wilkins_Nature_1996_phase_contrast_polychromatic_imaging,
*Chapman_NPhot_2010_Review_diff_imaging,
*Abbey_NPhot_2011_Broadband_Coherent_Diffractive_Imaging}.


In conclusion, irradiating gas jets with multi-terawatt lasers we observe
comb-like harmonic spectra reaching the `water window' region.
%
We propose the harmonic generation mechanism based on self-focusing,
nonlinear wave 
formation with electron density spikes, and collective
radiation by a compact electric charge. 
Catastrophe theory explains the electron spike sharpness and stability.
%
The maximum harmonic order scaling with the laser intensity 
will allow
reaching the keV range.
Our 
results 
open the way to a compact coherent X-ray source 
built on a university laboratory scale repetitive laser and accessible,
replenishable and debris-free gas jet target.
This will impact many areas requiring a bright X-ray/XUV source for pumping, probing, imaging, or attosecond science.

We acknowledge the financial support from MEXT (Kakenhi 20244065, 21604008, 21740302, and 23740413), JAEA President Grant, and STFC facility access fund.

\providecommand{\noopsort}[1]{}\providecommand{\singleletter}[1]{#1}%

\providecommand{\noopsort}[1]{}\providecommand{\singleletter}[1]{#1}%

\end{document}